\documentclass[twocolumn,showpacs,preprintnumbers,aps,floatfix]{revtex4-1}
\pdfoutput=1

\usepackage{amssymb,amsmath,graphicx,bm,textcomp,color}
\usepackage[squaren,Gray]{SIunits}

\renewcommand{\vec}[1]{\mathbf{#1}}
\newcommand{\moy}[1]{\ensuremath{\langle#1\rangle}}

\usepackage[normalem]{ulem}

\begin{document}

\title{Plastic flow and localization in an amorphous material: experimental interpretation of the fluidity}

\author{David Houdoux}
\author{Thai Binh Nguyen}
\author{Axelle Amon} \email{axelle.amon@univ-rennes1.fr}
\author{J\'er\^ome Crassous}

\affiliation{Univ Rennes, CNRS, IPR (Institut de Physique de Rennes) -
  UMR 6251, F-35000 Rennes, France}

\date{\today}

\begin{abstract}
{We present a thorough study of the plastic response of a granular
  material progressively loaded. We study experimentally the evolution
  of the plastic field from a homogeneous one to an heterogeneous one
  and its fluctuations in term of incremental strain. We show that the
  plastic field can be decomposed in two components evolving on two
  decoupled strain increment scales. We argue that the slowly varying
  part of the field can be identified to the so-called fluidity field
  introduced recently to interpret the rheological behavior of
  amorphous materials. This fluidity field progressively concentrates
  along a macroscopic direction corresponding to the Mohr-Coulomb
  angle.}
\end{abstract}

\maketitle

\section{Introduction}\label{sec:intro}
\begin{figure}[htbp]
\centering
\includegraphics[width=\linewidth]{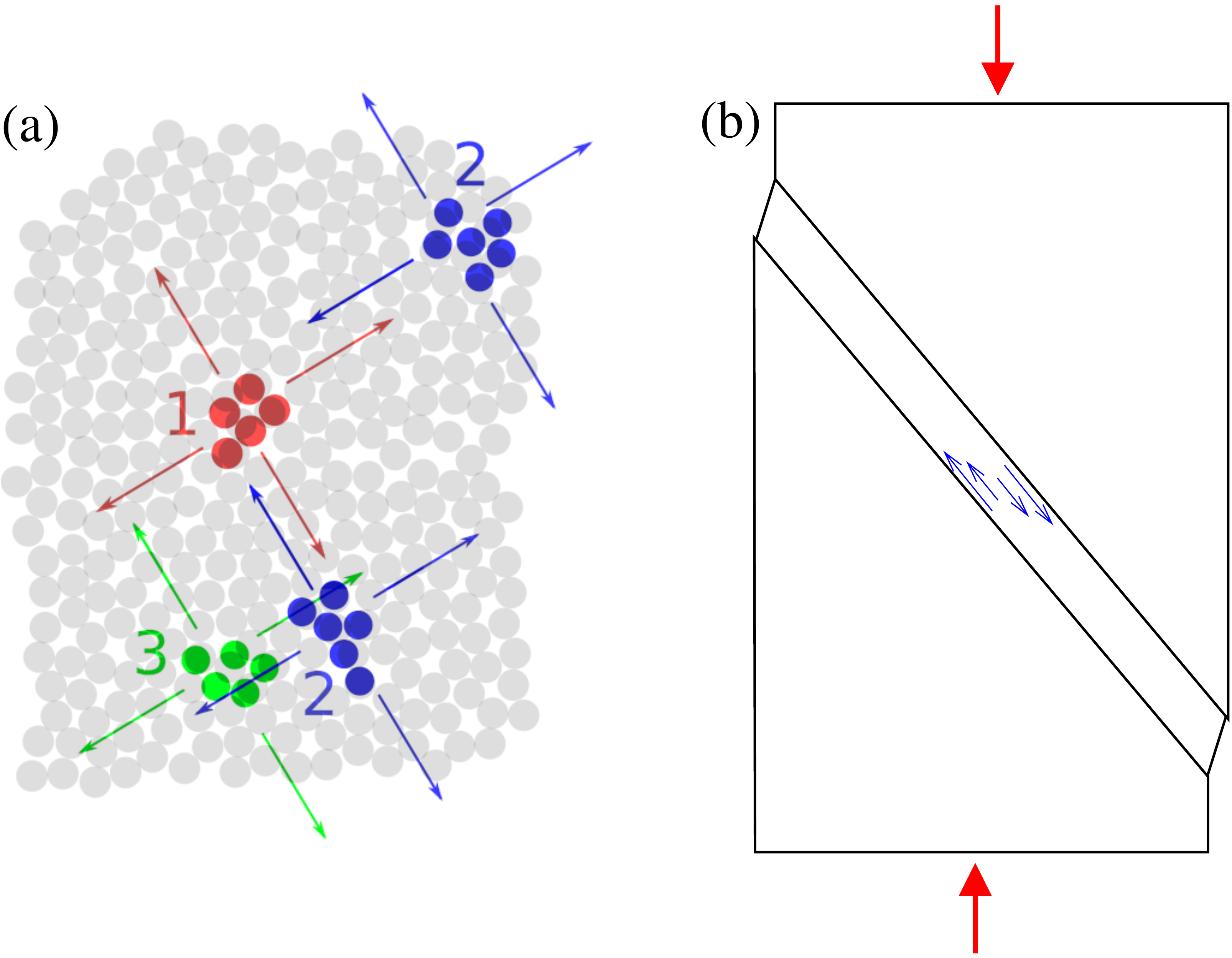}
\caption{Schematic pictures of a plastic flow in athermal
  materials. (a) At a microscopic scale, a small goup of elements
  reorganizes and redistributes mechanical stress in the
  material. This initial event can trigger a cascade of other
  rearrangements represented schematically with other colors to
  indicate their order of occurence. (b) At a macrocsopic scale, the
  flow concentrates into a narrow zone of the material such as a shear
  band.}
\label{fig:intro1}
\end{figure}

A physical description of the elementary mechanisms underlying the
plasticity of amorphous materials has emerged in the past few years
based on experimental evidences and numerical
investigations~\cite{Barrat2011}. One of the basic ingredient is the
fact that at an elementary level, plasticity occurs through local
plastic events implying only a few number of
constituents~\cite{Argon1979,Spaepen1977}, typically a few
tens~\cite{Kabla2003,Schall2007,Amon2012}. Such events are
schematically represented in Figure~\ref{fig:intro1}(a). When such
events occur, they redistribute
stress~\cite{Maloney2006,Tanguy2006,Tsamados2008}. This redistribution
can trigger other rearrangements generating avalanches of events. An
important point of this description is its universality as the stress
redistribution is independent of the local interaction between the
constituents. The mechanical properties that intervene in this
description are the elastic properties which characterize the
amorphous material on a large scale as an effective medium. The
rearrangement can be treated theoretically as a small inclusion in an
elastic matrix and the stress redistributed can be computed as shown
by J. D. Eshelby~\cite{Eshelby1957}. The coupling between the plastic
events is then quadrupolar.

While there is now an agreement about the validity of this picture in
the community, the question of how this microscopic picture builds up
in a macroscopic flow is still open.

For athermal materials, an important point is to model how the flow
can be self-sustained. A possible approach is at a coarse-grained
level. In order to describe rheology of yield stress material, Hebraud
and Lequeux~\cite{Hebraud1998} introduced a local internal variable
which represents the spatial and temporal density of plastic events
occurring at a given time and at a given position. At a given point,
this quantity evolves due to some stress relaxation, macroscopic
loading and a background of mechanical noise due to plastic events
occurring everywhere in this sample. The use of this internal
variable, usually called \emph{fluidity}, have been very fruitful to
explain the rheology of yield-stress fluid. However, numerous studies
evidence that \emph{local} constitutive equations for the rheology are
not compatible with the observations~\cite{Goyon2008}. A plastic flow
in a point of a material has an effect on the response of the material
at some distance.  A new class of \emph{nonlocal} constitutive models
have then been introduced~\cite{Goyon2008,Bocquet2009}. The length
characterizing the range of this interaction between events is called
the \emph{cooperativity length} and is supposed to depend on the
distance of the local stress to the yield stress. Such a model has been
adapted to granular materials and describes the stationary response of
granular materials in numerous
configurations~\cite{Kamrin2012,Henann2013}.

An important feature usually observed in amorphous materials is the
fact that strain localization are observed at a large
scale~\cite{Schall2010}. When a yield stress material is sheared
homogeneously the deformation is not homogeneously allocated in the
sample but is concentrated in thin parts, called shear bands, in which
the strain rate is large while the other parts of the material
experience small strain rate (Fig.~\ref{fig:intro1}(b)). A description
of how those stationary shear bands emerge from microscopic elementary
plastic events or from an homogeneous fluidity field is still
missing. At a macroscopic point of view, to describe shear bands is to
consider a stress-based failure criterion. When the local stress is
larger than a threshold, called the \emph{yield stress}, the material
flows. In the particular case of granular materials, the failure
criterion is given by the \emph{Coulomb} threshold: failure occurs
when the ratio of the local shear stress to the pressure is larger
than $\tan \phi$ where $\phi$ is the \emph{angle of internal
  friction}. Experimentally the internal friction angle is defined
from the value of the yield stress. It may also been determined
from the angle between the failure plane and the principal stress
direction using a \emph{Mohr-Coulomb} construction. Such a description
of plasticity of shear bands has several limitations and in particular
the fact that no lengthscale is introduced in the failure criterion so
that the width of the shear band cannot be predicted.

In summary, the picture emerging from the literature is the
following. At a microscopic scale the particles move. At a mesoscopic
scale, individual plastic events may be defined, and such events are
coupled by elasticity. At a macroscopic scale, the rate of such events
may be theoretically represented by a variable called fluidity, which
varies at the scale of the flow. Finally, macroscopic experimental
observations show that strain localization occurs. Experiments or
numerical simulations showing {\it simultaneously} those different
behaviors are missing. We propose in this study an experiment that
evidences many features of those plastic flow behaviors. For this we
performed experiments on a shear flows of athermal spheres. Using an
interferometric technique, we are able to follow the fluctuations of
plasticity. Those fluctuations evidence some features of individual
plastic events, such as the coupling of events by elasticity. If those
fluctuations are averaged, a slowly varying field of deformation
emerges, that may be identified with the fluidity field defined
theoretically. Depending of the temporal scales at which the
plasticity field is observed, different behaviors at mesoscopic and
macroscopic scales may be evidenced simultaneously.

The manuscript is organized in the following way. In
section~\ref{sec:setup} we present the experimental setup. In
section~\ref{sec:correl}, the behavior of the correlation functions of
the plasticity field is investigated. Those correlations functions can
be separated into a slow and a fast component as shown in
section~\ref{sec:study}. Finally, in section~\ref{sec:discussion}, we
discuss the identification of the slow component of the plastic flow
to the fluidity field.

\section{Experimental Set-up}\label{sec:setup}

\begin{figure}[htbp]
\includegraphics[width=\columnwidth]{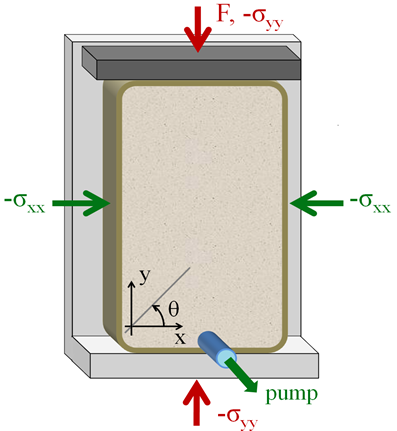}
\caption{Schematic representation of a biaxial setup. The granular
  material is enclosed between a latex membrane and a glass plate (not
  represented here). A partial vacuum inside the sample creates a
  confining stress $-\sigma_{xx}=30$~kPa. At the top, a moving plate
  exerts a compression of the sample at fixed velocity along the
  $y$-axis. The back plate as well as the front glass one forbid
  displacement along the $z$-direction ensuring plane-strain
  conditions.}
\label{fig:biaxial}
\end{figure}

\begin{figure}[htbp]
	\includegraphics[width=\columnwidth]{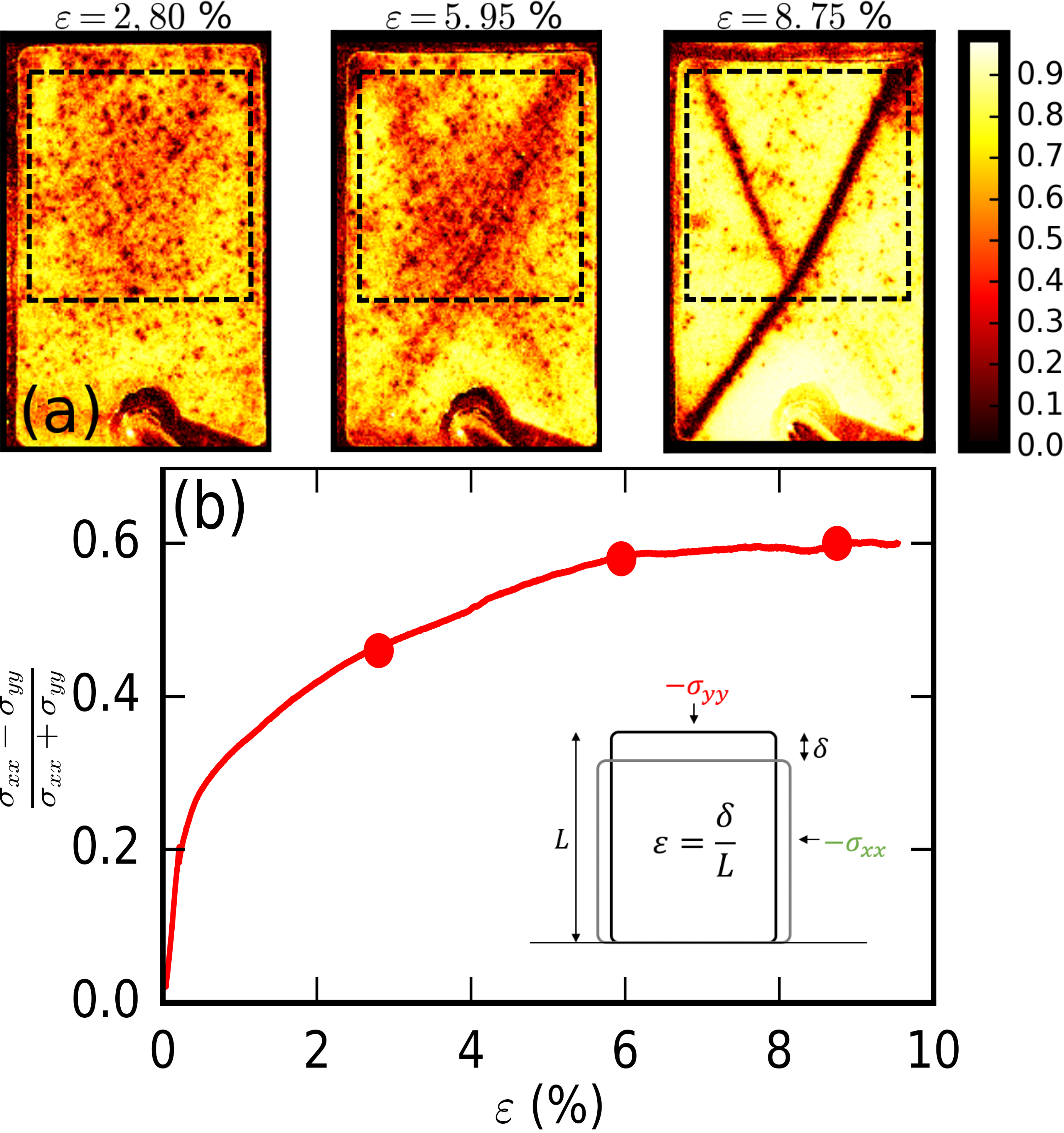}
	\caption{ (a) Correlation maps $g_I$ for three different
          values of $\varepsilon$ (2.80 \%; 5.95 \% and 8.75 \%). The
          region of interest used for image analysis is shown in
          dashed line. (b) Loading curve. Dots symbolize the positions
          of the three correlation maps. Inset : notations.}
	\label{fig:gI}
\end{figure}

The experimental setup consists of a biaxial compressive test in plane
strain conditions already described extensively in
~\cite{LeBouil2014a}. This kind of plane-strain compression is also
very common in geomechanics ~\cite{Viggiani2004}. The granular
material (dry glass beads of diameter \unit{d=70-110}{\micro\meter},
initial volume fraction $\approx 0.60$) is placed between a preformed
latex membrane (\unit{85 \times 55 \times 25}{\milli\meter\cubed}) and
a glass plate. A pump produces a partial vacuum inside the membrane,
creating a confining stress \unit{-\sigma_{xx}\simeq 30}{\kilo\pascal}
(see Fig.~\ref{fig:biaxial}). The sample is placed in the biaxial
apparatus where displacement normal to the $xy$ plane is prevented by
the front glass plate and a back metallic one, ensuring plane-strain
conditions. At the bottom the sample is blocked by another metallic
plate, while an upper plate is displaced vertically by a stepper
motor.  The stress applied at the top of the sample is then
$-\sigma_{yy}=-\sigma_{xx}+F/S$, where $F$ is the force measured by a
sensor fixed to the upper plate, and S the section of the sample. The
velocity of the motor is
\unit{1}{\micro\meter\cdot\reciprocal\second}, leading to a
deformation rate \unit{\dot{\varepsilon}=1.2 \times
  10^{-5}}{\reciprocal\second} (where the deformation is defined as
$\varepsilon=-\varepsilon_{yy}=\delta/L$ - see inset
Fig.~\ref{fig:gI}). The corresponding loading curve is presented in
Fig.~\ref{fig:gI}(b). We work in the quasistatic regime. Note that the
stress gradient due to gravity is negligible, and the value of the
confining stress is too low to expect a crushing of particule. To
break the symmetry when failure occurs, for instance, to study the
behavior of the shear band (see~\cite{nguyen2016}), the metallic plate
at the bottom of the sample can freely translate in the x-direction
thanks to a roller bearing. In the following, experiments that are
presented are made in this latter configuration.

Deformations are observed through the front glass plate using a
Diffusing Wave Spectroscopy (DWS) method already described before in
~\cite{erpelding2008,erpelding2013}. A laser beam at
\unit{532}{\nano\meter} is expanded to illuminate the entire
sample. The light undergoes multiple scattering inside the granular
material and we collect the backscattered rays. The latter interfere
and form a speckle pattern. The image of the front side of the sample
is recorded by a $7360\times4912$ pixels camera. Speckle images are
subdivised in square zones of size $16\times16$ pixels and compared
using a correlation method explained elsewhere
~\cite{erpelding2008,Amon2017}. For each zone the correlation between
two successive images 1 and 2 is computed as follows:

\begin{equation}\label{eq:gI}
g_I^{(1,2)}=\frac{\moy{I_1I_2}-\moy{I_1}\moy{I_2}}{\sqrt{\moy{I_1^2}-\moy{I_1}^2} \sqrt{\moy{I_2^2}-\moy{I_2}^2}}
\end{equation}

\noindent where $I_1$ and $I_2$ are intensities of the pixels of two
successive images and the averages $\moy{.}$ are done over the
$16\times16$ pixels of a zone. We obtain correlation maps where each
pixel is calculated from eq.~\eqref{eq:gI} and corresponds to a volume
of area in the front plane $4d\times4d$ and depth of a few
$d$. Examples of such maps of correlation are given in
Fig.~\ref{fig:gI}(a). The normalisation used in (\ref{eq:gI}) leads to
values for $g_I$ in the interval $[0,\, 1]$ (see the colorscale). The
decorrelation of the backscattered light (i.e. low value of $g_I$)
comes from relative beads motions as, for example, combinations of
affine and nonaffine bead deplacements or rotation of nonspherical
beads. In the following, maps of correlation are calculated between
two successive images with a fixed axial deformation increment
$\delta\varepsilon=3.5\times10^{-5}$.

\section{Spatio-Temporal correlation functions}\label{sec:correl}

Observations made on the correlation maps during the loading are the
same as those already presented in
~\cite{LeBouil2014a,LeBouil2014b,nguyen2016}. In the sample,
deformations are distributed inhomogeneously (see
Fig.~\ref{fig:gI}(a)). At the beginning of the loading, plastics
events are randomly distributed throughout the sample. After a
deformation of a few percent ($\varepsilon\simeq3\,\%$) these plastic
activities are organized along intermittent micro-bands until one or
two permanent shear bands are established. We relate the former to
avalanches of local rearrangements predicted in the microscopic
descriptions of the plasticity of amorphous materials. The latter are
in agreement with the Mohr-Coulomb model. In the following, we present
a method to isolate each of these contributions to plasticity.

\subsection{Spatio-temporal correlation function}\label{subsec:STCfunction}

First, we introduce a new parameter
$a(\vec{r},\varepsilon)=1-g_I^{(1,2)}$ where $g_I^{(1,2)}$
(see~(\ref{eq:gI})) is calculated on a square area of ($4d\times4d$)
at position $\vec{r}$. $I_1$ and $I_2$ are intensities of pixel images
taken respectively at $\varepsilon_1=\varepsilon$ and
$\varepsilon_2=\varepsilon+\delta\varepsilon$. Thus
$a(\vec{r},\varepsilon)$ represents the activity at position $\vec{r}$
around the deformation $\varepsilon$.

We define also a spatio-temporal correlation function as follow :
\begin{multline}\label{eq:C}
C(\varepsilon, d\varepsilon, d\vec{r})=\moy{a(\vec{r'}, \varepsilon')\cdot a(\vec{r'}+d\vec{r},\varepsilon'+d\varepsilon)}\\
-\moy{a(\vec{r'}, \varepsilon')}\moy{a(\vec{r'}+d\vec{r},\varepsilon'+d\varepsilon)}
\end{multline}

\noindent where $\vec{r'}$ and $\vec{r'}+d\vec{r}$ have to be in the
region of interest (ROI, see Fig.~\ref{fig:gI}(a)) and the deformation
$\varepsilon'$ in the interval
$[\varepsilon;\varepsilon+\Delta\varepsilon]$. Keep in mind that $g_I^{(1,2)}$
and consequently $a(\vec{r},\varepsilon)$ are calculated over an increment
$\delta\varepsilon=3.5\times10^{-5}$. So there is no integration of mean displacement
over $[\varepsilon\,;\,\varepsilon+\Delta\varepsilon]$. Averages are made over
all pairs of pixels for which $\vec{r'}$ and $\vec{r'}+d\vec{r}$ are
in the ROI and over a deformation span $\Delta\varepsilon$. In the
following, the value of $\Delta\varepsilon$ is fixed at
$\Delta\varepsilon=0.35\,\%$. This value is large enough to have good
averages but remains small compare to the total deformation
applied. Note that in all the following the deformation plays the role
of time. Our experiments are in quasi-static conditions. A process
which can be observed only for small deformation increments will be
called a \emph{fast} process in the following while a feature that
lasts for large strain increments will be called a \emph{slow}
process.

\subsection{Temporal scale of fluctuations}\label{subsec:temporal}

Throughout the loading, we can observe on the correlation maps that
micro-bands are fluctuating rapidly while shear bands evolves slowly. To
quantify the temporal scale of the fluctuating part, we compute
temporal correlations as follow:
\begin{equation}\label{eq:CT}
C_T(\varepsilon, d\varepsilon)=C(\varepsilon, d\varepsilon, d\vec{r}=\vec{0})
\end{equation}

\noindent where $C_T$ can be interpreted as a spatial autocorrelation
function giving only the temporal information.

\begin{figure}[!h]\centering
\includegraphics[width=\columnwidth]{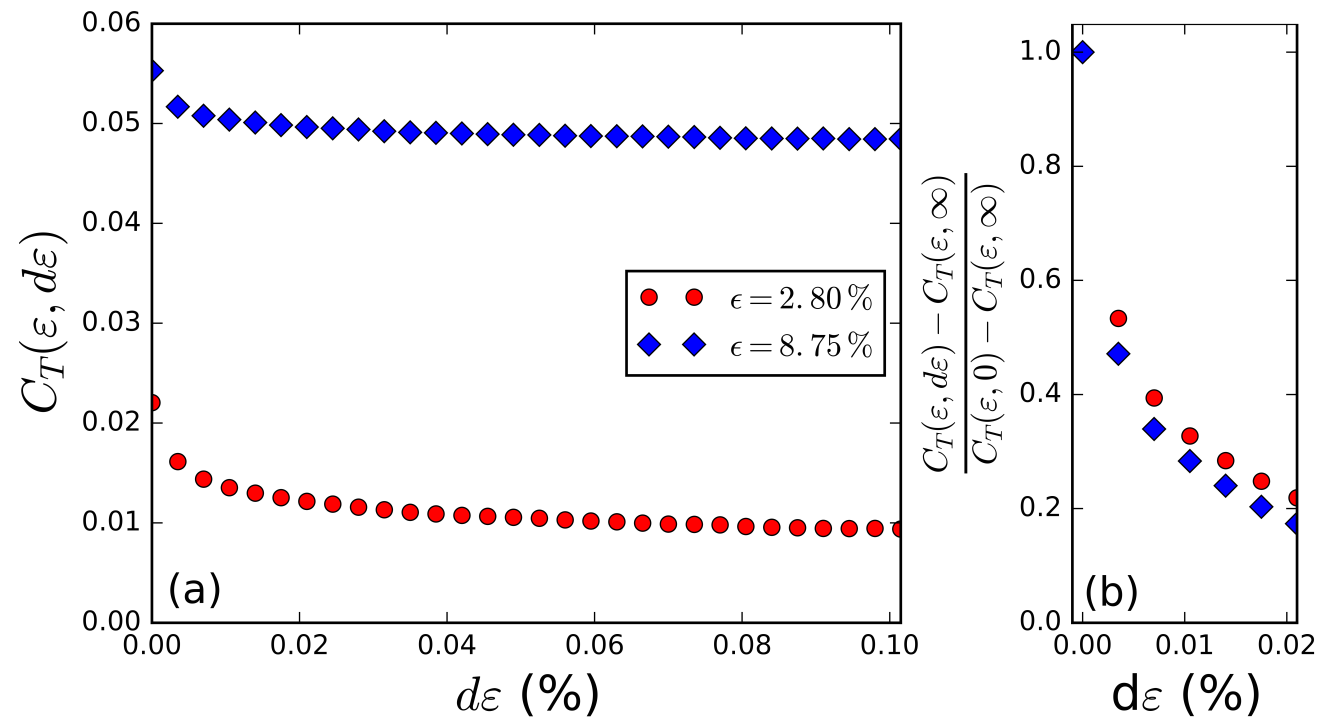}
\caption{(a) Temporal correlation function as a function of
  $d\varepsilon$ at the beginning ($\varepsilon=2.80\%$) and at the
  end ($\varepsilon=8.75\%$) of the loading. (b) Rescaled correlation
  functions showing the initial drop over a characteristic gap
  $d\varepsilon^*\approx0.01\,\%$.}
\label{fig:CT}
\end{figure}

Fig.~\ref{fig:CT} shows $C_T(\varepsilon, d\varepsilon)$ as a function
of $d\varepsilon$ at two points in the loading $\varepsilon=2.80\,\%$
and $\varepsilon=8.75\,\%$. A rapid decorrelation for a gap
$d\varepsilon^*\approx0.01\,\%$ is initially observed, followed by a
second smoother decay which tends to a non-zero finite value because
of the heterogeneity of the mean field. The scale in deformation
increment over which the initial drop takes place,
$d\varepsilon^*\approx0.01\,\%$, gives us an estimation of the
temporal scale of the fluctuating part.

\subsection{Separation method}\label{subsec:separation}

The aim is to separate the two plastic phenomena throughout the
loading. Our method is based on their different temporal
properties. As discussed in section~\ref{subsec:temporal}, the
fluctuating part is no longer correlated for a gap of deformation
larger than $d\varepsilon^*$. In the case when $d\varepsilon\gg
d\varepsilon^*$, the spatio-temporal correlation function
$C(\varepsilon, d\varepsilon, d\vec{r})$ corresponds only to the slow
part. In the following, we define the correlation function for the
slow part by taking arbitrary
${d\varepsilon=\Delta\varepsilon/2=0.175\,\%\gg d\varepsilon^*}$ for
the spatio-temporal correlation function:
\begin{equation}\label{eq:Cslow}
C_\text{slow}(\varepsilon, d\vec{r})=C\left(\varepsilon, \frac{\Delta\varepsilon}{2},d\vec{r}\right)
\end{equation}

In the same way, for $d\varepsilon\ll d\varepsilon^*$ the
spatio-temporal correlation function $C(\varepsilon, d\varepsilon,
d\vec{r})$ contains all the fluctuations. Thus, we define the total
correlation function by taking $d\varepsilon=0$ as follows:
\begin{equation}\label{eq:Ctot}
C_\text{tot}(\varepsilon,d\vec{r})=C(\varepsilon,0,d\vec{r})
\end{equation}

Finally, the correlation function describing the fluctuating part is
obtained as follows:
\begin{equation}\label{eq:Cfluc}
C_\text{fast}(\varepsilon, d\vec{r})=C_\text{tot}(\varepsilon,
d\vec{r})-C_\text{slow}(\varepsilon, d\vec{r})
\end{equation}

\begin{figure}[!h]
\includegraphics[width=\columnwidth]{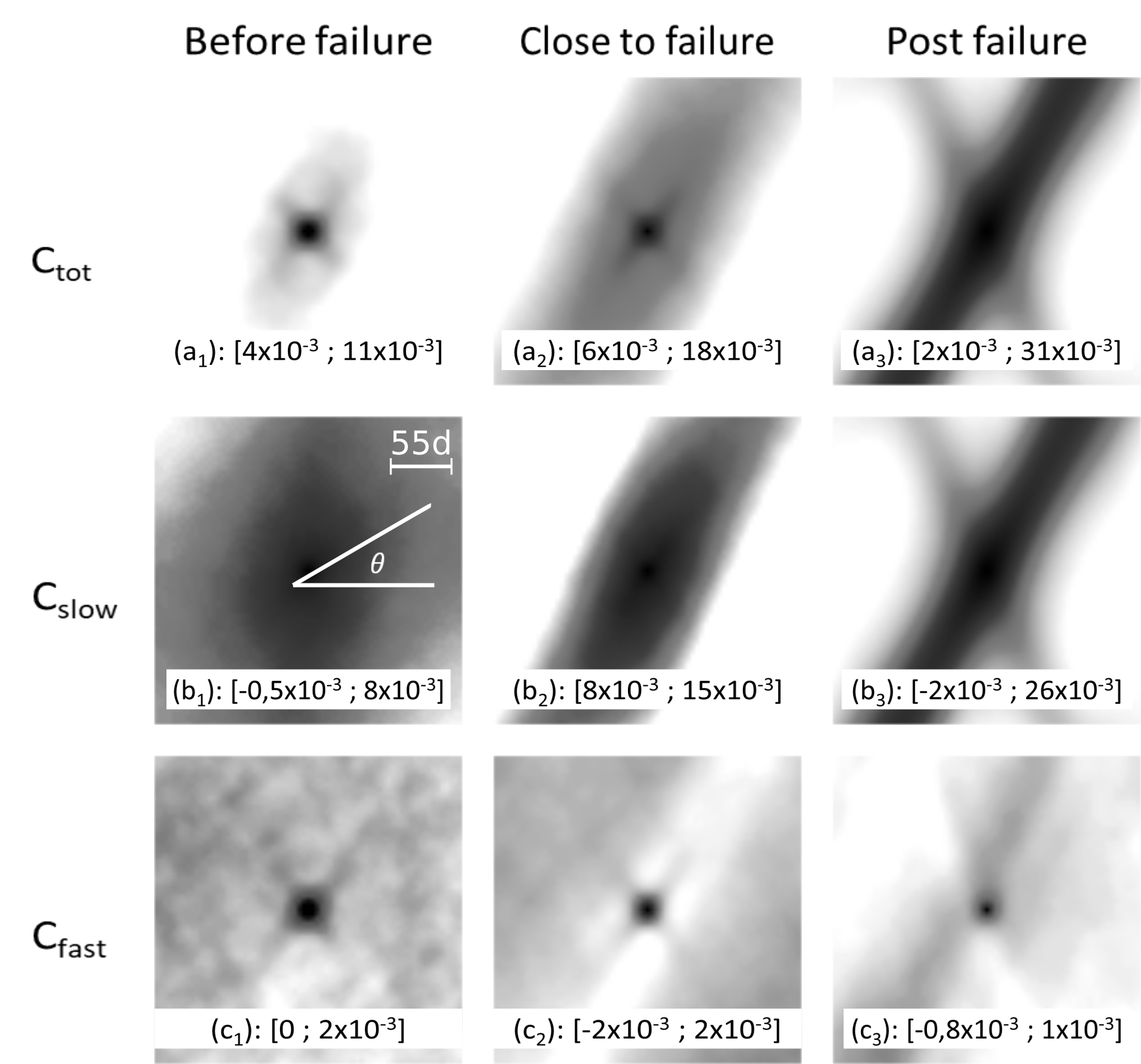}
\caption{Spatial representation of the different parts of the
  correlation function for different values of the loading. All maps
  are centered at $\protect d\vec{r}=\protect\vec{0}$. The spatial
  scale, in unit of beads diameter, is represented in (b1). Top row
  ((a1), (a2) and (a3)): total correlation function $C_\text{tot}$;
  Middle row ((b1), (b2) and (b3)): slow part $C_\text{slow}$ of the
  correlation function; Bottom row ((c1), (c2) and (c3)): fast part
  $C_\text{fast}$ of the correlation function. Left column, the system
  is well before failure, $\varepsilon=2.80\,\%$; Middle column, it is
  close to the failure $\varepsilon=5.95\,\%$; Right column, after
  failure $\varepsilon=8.75\,\%$. Scale represented by intervals corresponds to white and black pixels respectively.} \label{fig:separation}
\end{figure}

Fig.~\ref{fig:separation} shows the correlations functions
$C_\text{tot}$, $C_\text{slow}$ and $C_\text{fast}$ defined
previously. Those correlations functions are computed at different
macroscopic deformation $\varepsilon=2.80\,\%$ (well before failure),
$\varepsilon=5.95\,\%$ (close to failure), and $\varepsilon=8.75\,\%$
(post-failure).

\section{Experimental results}\label{sec:study}

\subsection{Behavior of the slow and fast parts}\label{subsec:behavior}

In this subsection, we discuss the results obtained by the calculation
of the slow and fast parts of the correlation functions, and we
evidence the differences of behavior between the long and the short
time correlation functions.

We first discuss the correlation function near failure (middle column
in Fig.~\ref{fig:separation}) where the characteristic behavior is
clearly visible. The total correlation function $C_\text{tot}$ (see
Fig.~\ref{fig:separation}(a2)) displays two different spatial
features. Near $d{\bf r}=0$, we observe a high correlation along two
directions, forming a small cross at the center of the figure. At
large distance, the correlation prevails along a large inclined band
spanning the full image. The fast part and the slow part of the total
correlation function displays each only one of those two spatial
behaviors. The fast part displays only the small cross near $d{\bf
  r}=0$ (Fig.~\ref{fig:separation}(c2)), whereas the large inclined
band belongs to the slow part (Fig.~\ref{fig:separation}(b2)).

In the pre-failure stage, Fig.~\ref{fig:separation}(c1) shows a
fluctuating contribution at short distances, but the slow part of the
correlation function (Fig.~\ref{fig:separation}(b1)) remains spread on
the full image and displays no structure. After failure, the slow part
of the correlation function appears clearly correlated along two bands
(Fig.~\ref{fig:separation}(b3)), whereas the quadrupolar structure of
the fluctuating part near the origin is no longer visible.

\subsection{Geometrical characterisation of the correlations.}

Fig.~\ref{fig:separation} evinces that the fast and slow components of
the correlation functions are both anisotropic, but with slightly
different directions. We now use this observation to quantify the
duality of the correlation functions with the time-scale.

\begin{figure}[!h]
\includegraphics[width=\columnwidth]{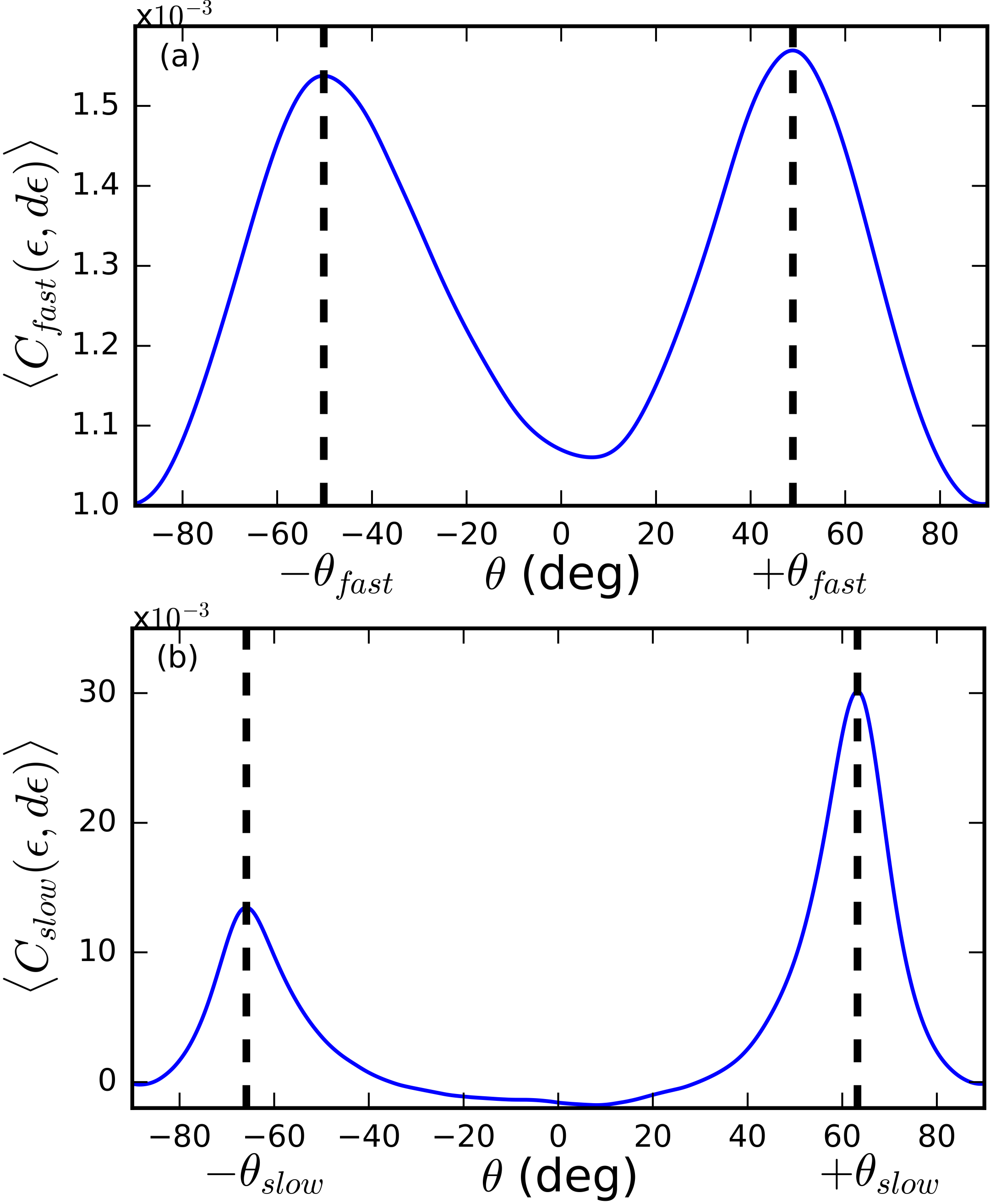}
\caption{Angular projection of the spatial correlation functions,
  i.e. mean value of the pixels intersected by a line passing through
  the center of a spatial correlation function and making an angle
  $\theta$ with the $x$-axis as a function of $\theta$ for (a)
  $C_\text{fast}$ at $\varepsilon=5.95\,\%$ and (b) $C_\text{slow}$ at
  $\varepsilon=8.75\,\%$.}
\label{fig:Proj}
\end{figure}

For this, we need to determine the angles at which the correlation
enhancement occurs. We use a projection method inspired by what has
been done previously~\cite{nguyen2016}. As the correlation maps are
centered (see Fig.~\ref{fig:separation}), we compute the mean value of
the pixels intersected by a line passing through the center and with
an angle $\theta$ with the $x$-axis. Next we plot these mean values as
a function of the angle $\theta$ between $-\pi/2$ and
$+\pi/2$. Fig.~\ref{fig:Proj}(a) shows the result of such projection
of the fast part of the correlation $C_\text{fast}$ for an axial
deformation $\varepsilon=5.95\,\%$. Fig.~\ref{fig:Proj}(b) shows the
projection of the slow part of the correlation $C_\text{slow}$ for
$\varepsilon=8.75\,\%$. From a projection profile as the ones obtained
in Fig.~\ref{fig:Proj}, we determine the angle characterizing a
response as the mean of the absolute values of the angles at which
occur the maxima of the curve. Error bars are determined as the
difference between the angles at which the maxima occur and the ones
obtained from gaussian interpolations of the projection profile on
each halves of the $[-\pi/2,\pi/2]$ interval.

\begin{figure}[!h]
	\includegraphics[width=\columnwidth]{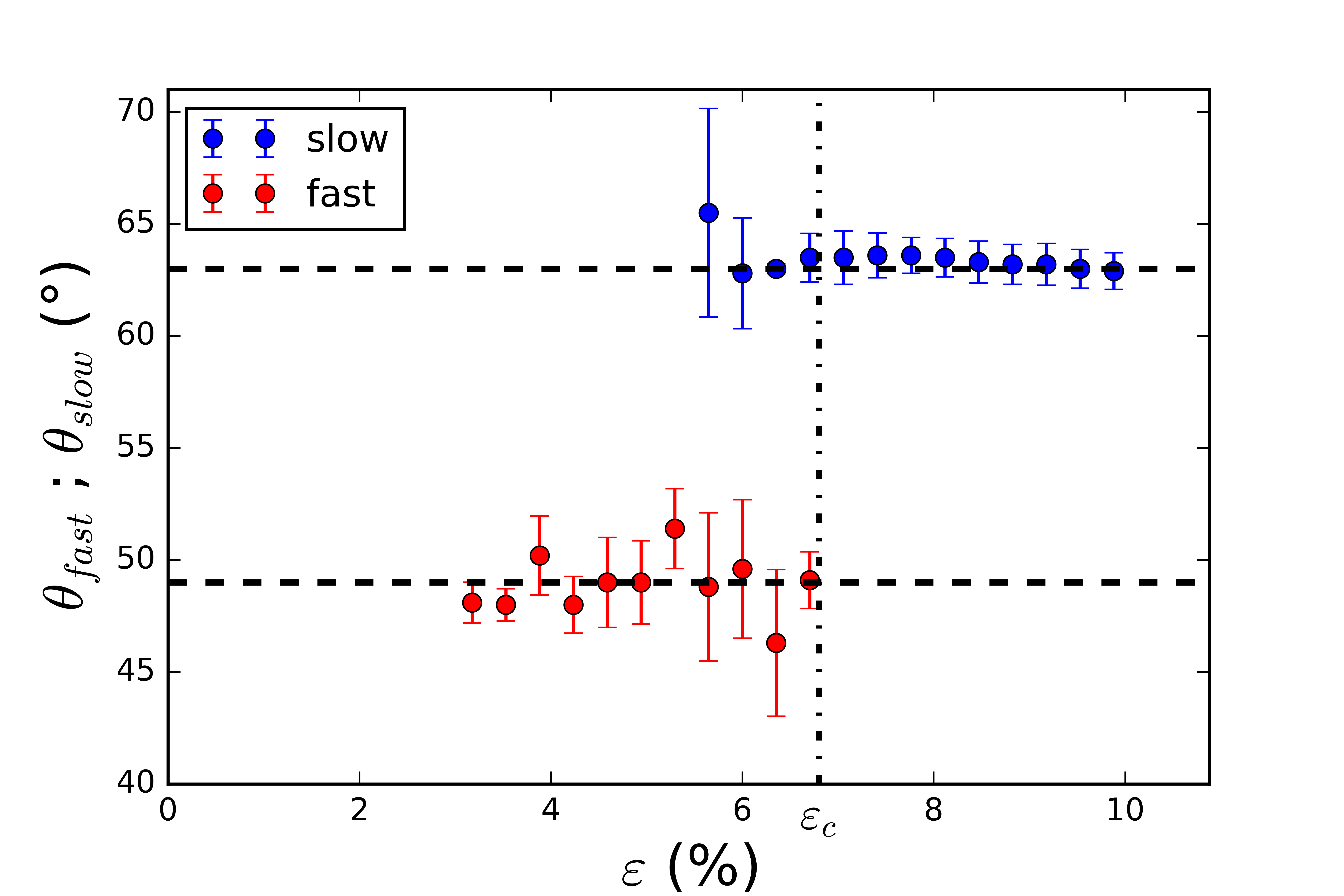}
	\caption{Angles of the fast and the slow parts of the spatial
          correlation function. A coexistence of two distinct
          directions is observed just before the failure represented
          by the dashed line. $\varepsilon_c$ stands for the critical
          value of the strain when the failure occurs.}
	\label{fig:angles}
\end{figure}

Fig.~\ref{fig:angles} shows the values of the angles at which the fast
and the slow part of the correlation function are respectively
maximals as a function of the macroscopic deformation in the
$\varepsilon\in[2.45\,\%;9.80\,\%]$ range. The failure occurs for
$\varepsilon\simeq6.65\,\%$. We observe that angles at which
correlation occurs are clearly different for the fluctuating and the
slow parts. For a deformation in the range
$\varepsilon\in[5.60\,\%;6.65\,\%]$ the coexistence of a fast and slow
correlation oriented at different angles is clearly visible. Outside
this range of deformation, we were not able to define two different
angles simultaneously in the correlation functions.

\section{Discussion}\label{sec:discussion}

\subsection{Interpretation of geometrical angles for slow and fast
  correlation functions.}

\begin{figure}[h!]
\centering
\includegraphics[width=\columnwidth]{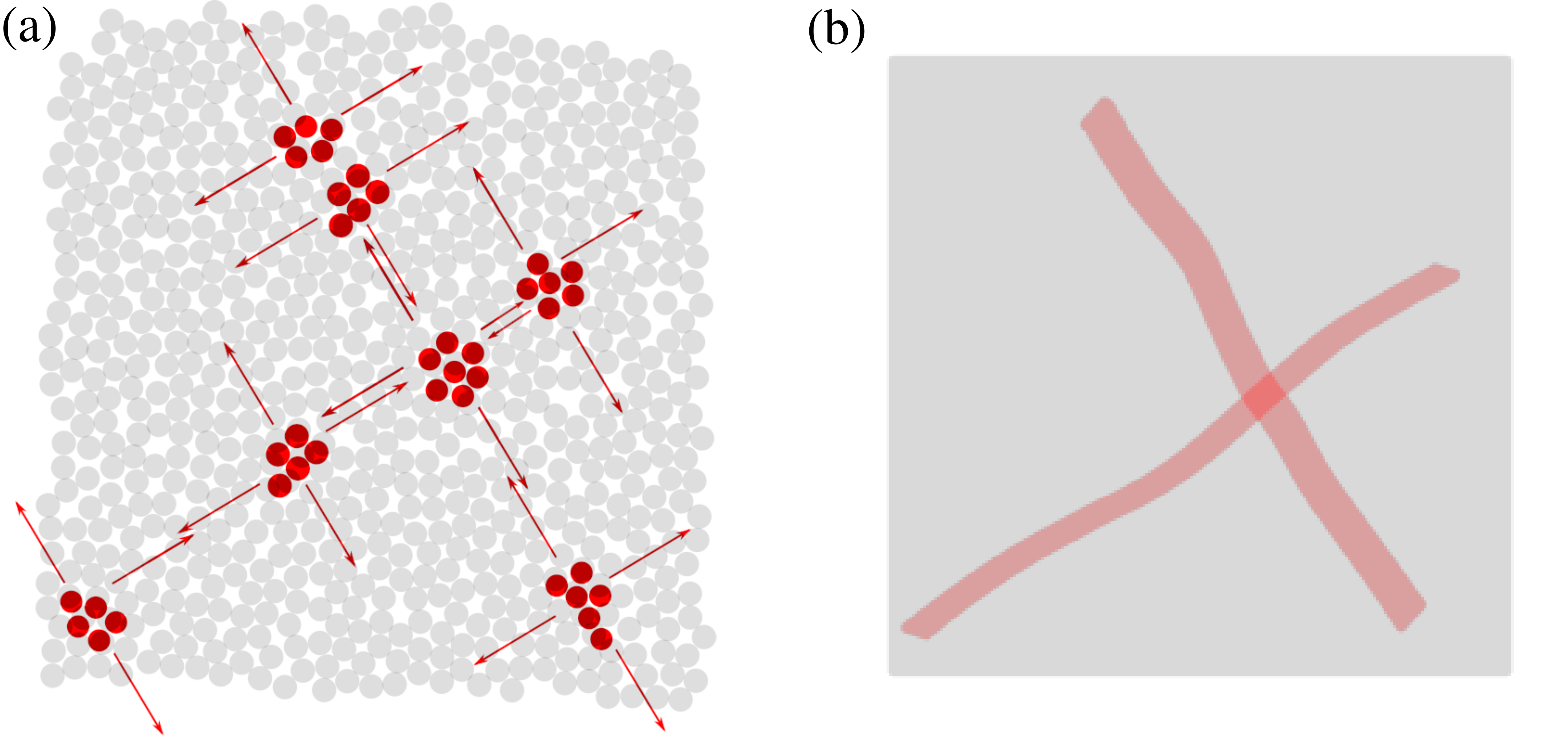}
\caption{Schematic representation of our interpretation of the
  microbands as cascades of elementary events. (a) Local
  rearrangements coupled by elasticity. (b) Resulting deformation
  integrated on a larger time-scale.}
\label{fig:detail}
\end{figure}

As it has been shown previously, we clearly observe two different
angles for the correlation function. The fast part of the correlation
function is quadrupolar, with an average inclination $\theta_E \simeq
48^{\circ}$. This orientation may be directly linked to the direction
of maximum stress redistribution given by the Eshelby tensor. We can
interpret the fast correlation of the plastic events as small cascades
of events as represented schematically in
Figure~\ref{fig:detail}. Such fluctuations has also been reported in
few numerical studies~\cite{Kuhn1999,Maloney2006}.

The slowly varying part of the correlation function is oriented at an
angle $\theta_{MC} \simeq 63^{\circ}$. This orientation may be related
to the formation of a macroscopic shear band into the material. The
orientation of the shear band depends on the value of internal
friction ratio accordingly to a Mohr-Coulomb construction. This slow
variation is related to large scale cooperative effects.  It is
important to note that two scales, and the two orientations coexist
during the approach to failure (see Fig.~\ref{fig:angles}): the
fluctuations of plasticity are oriented along $\theta_E$ whereas the
mean plasticity is oriented along $\theta_{MC}$. The two angles seem
relatively constant during the loading process, and no intermediate
orientations are observed, even close to failure.

\subsection{Schematic interpretation of the experiment.}
\begin{figure*}[p]
\centering
\includegraphics[width=\textwidth]{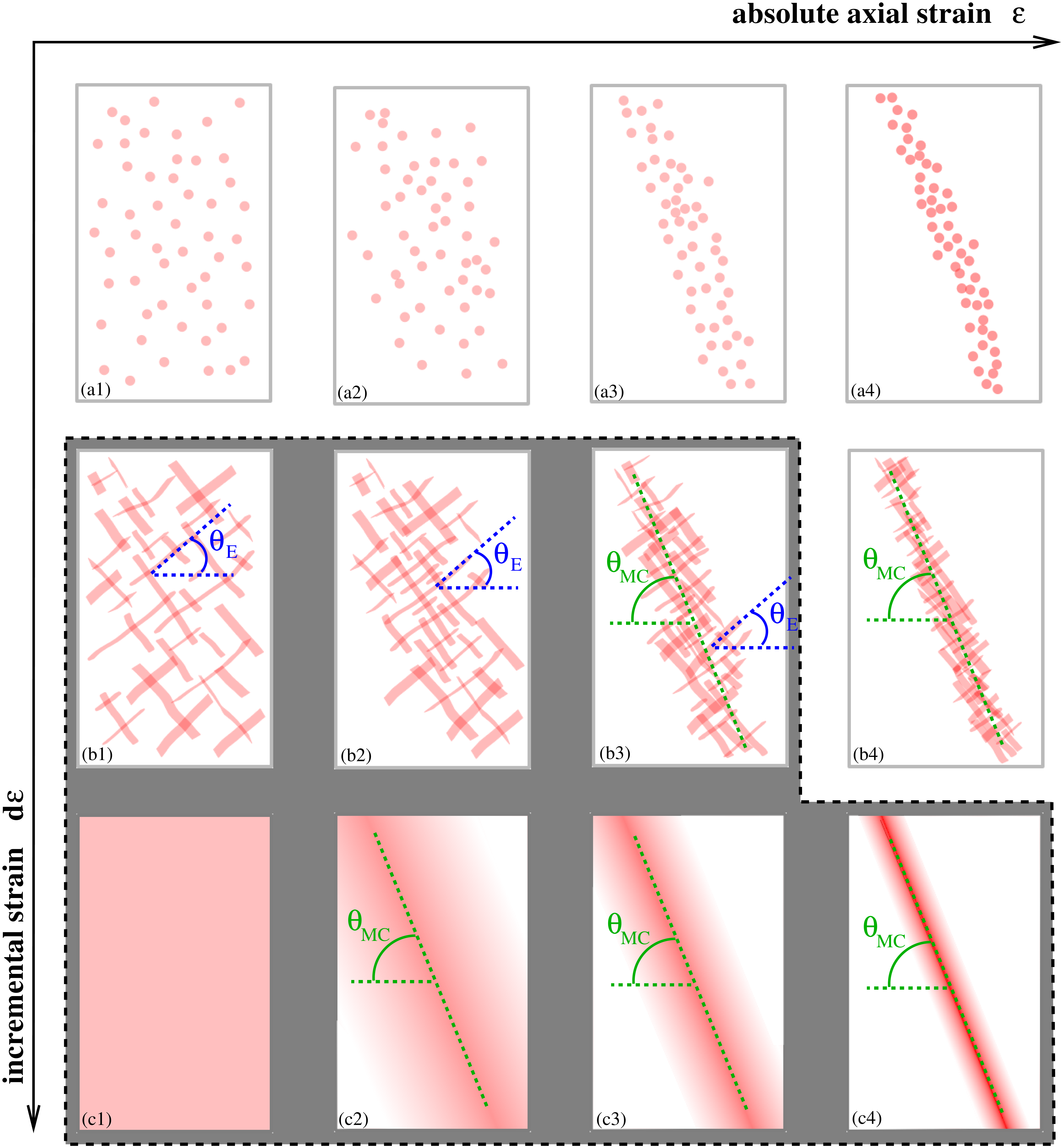}
\caption{Schematic representation of the plastic activity at different
  values of the axial loading $\varepsilon$ (horizontal axis) and for
  different strain increments $d \varepsilon$. At an elementary
  level, plasticity is awaited to occur through local plastic events
  represented as dots on the first line. When integrating those spots
  on a larger strain increment, a correlated micro-strucutre emerges
  governed by the elastic properties of the bulk material (second
  line). The angle characterizing this elastic coupling, $\theta_E$,
  is given by the Eshelby kernel. On a larger strain increment this
  fluctuating pattern is lost and only a slowly evolving
  coarse-grained field remains. Shear band formation can be observed
  on this slow field as the emergence of an orientation $\theta_{MC}$
  in the mean plastic field. The grey area displays the features
  actually observed experimentally. The reason why (b4) is not
  observed experimentally is linked to the nonlinearity in deformation
  of the measurement methods: the observed field saturates in the
  band. (a1)-(a4) are not observed because the acquisition rate is too
  small to reach small enough $d \varepsilon$.}
\label{fig:discussion}
\end{figure*}

Figure~\ref{fig:discussion} summarizes schematically our results. At
the microscopic level, some grains reorganize when the applied stress
is increased: this is one "elementary event", that is represented as a
dot in Fig.~\ref{fig:discussion}(a1). This behavior can be observed at
the very beginning of the experiment. On a larger time-scale, we
observe correlated cascades of events which fluctuate rapidly
Fig.~\ref{fig:discussion}(b1). Those micro-avalanches are inclined
along a direction~$\theta_{E}$ given by the Eshelby stress tensor (see
Fig.~\ref{fig:detail}). Increasing further the deformation field on a
larger time scale, we observe a continuous field of plasticity (see
Fig.~\ref{fig:discussion}(a3)).

At the beginning of the loading, the events repartition is roughly
homogeneous at the sample scale (Fig.~\ref{fig:discussion}(a1)) and
the corresponding mean plastic field is also homogeneous
(Fig.~\ref{fig:discussion}(c1)). When the loading progresses, the
events mainly accumulate on an active zone
(Fig.~\ref{fig:discussion}(a2) and (a3)), and the plasticity field
then becomes heterogeneous (Fig.~\ref{fig:discussion}(c2) and
(c3)). This zone of activity is the precursor zone of the final shear
band, and is inclined at an angle~$\theta_{MC}$. Finally, at failure
(Fig.~\ref{fig:discussion}(c4)), all the activity is concentrated on a
narrow and stationary shear band.

\subsection{Fluidity as slow part of the plastic field.}
The physical interpretation of what is called fluidity is still a
debated issue~\cite{Bouzid2015}. The literature agrees on the fact
that it corresponds to a coarse-grained field directly linked to the
local plastic activity. Among the hypotheses underlying the Kinetic
Elasto-Plastic model of Bocquet et al.~\cite{Bocquet2009} an important
point is a Boltzmann Stosszahlansatz-like hypothesis of the decoupling
of the plastic-event dynamics. This means that the fluidity field is
defined on a time scale large enough for the memory of the coupling
between the underlying events to be lost.

We interpret the slow part of the plastic field
(Fig.~\ref{fig:discussion}(c1) to (c4)) as the field of the so-called
fluidity. Indeed, this field evolves smoothly spatially and has lost
the quadrupolar symmetry of the underlying fluctuations of the
plasticity. It corresponds to a coarse-grained field reflecting the
local plastic activity and thus matches the theoretical definition of
fluidity. 

In the specific case of sheared granular materials, it has been
proposed recently that a granular fluidity $g$ may be defined which is
proportional to a local shear rate $\dot \gamma$ and which is also
proportional to the velocity fluctuations $\delta
v$~\cite{Zhang2017}. Experimentally (see~\cite{Amon2017} for details),
the speckle decorrelation is a function of the affine and the
non-affine displacements of particles $\sqrt{\xi_0^2
  f(\varepsilon)+<\delta \xi^2>}$, where $\xi_0$ is a length of the
order of the grain diameter, $f(\varepsilon)$ is a quadratic function
of the affine deformation field $\varepsilon$, and $<\delta \xi^2>$ is
the mean square non-affine displacement. The spatial scale on which
the affine displacement field is defined is few bead
diameters~\cite{Amon2017}. For a sheared granular material, during a
time $\tau$, we expect that $f(\varepsilon)\sim (\dot \gamma
\tau)^2$~\cite{Amon2017} and $<\delta \xi^2> \sim (\delta
v~\tau)^2$. It follows that $\sqrt{\xi_0^2 f(\varepsilon)+<\delta
  \xi^2>} \propto g$. So our experimental probe of plastic activity
seems closely related to the above-mentionned definition of the
granular fluidity.

An important point in the definition of fluidity is the nature of the
coarse-graining at play to obtain a smooth field. Theoretically
fluidity is a variable obtained by a coarse-graining in space and
time. It is unclear in the literature if this last coarse-graining is
in ``true'' time or in strain increment. In our experiment, our
technique of measurement gives intrinsically a coarse-grained
measurement in space because it is based on multiple scattering of
light (see~\cite{erpelding2008,Amon2017} for a thorough discussion of
the spatial resolution of the method). The measurement is also
time-averaged as the acquisition rate of the camera is always much
smaller than the inverse of the time of propagation of sound waves in
the system. In addition to those resolution-based averages, we have
studied in the present article the effect of a coarse-graining in
strain increment on the observed plastic field. An important
conclusion drawn from our observations is the fact that for a large
enough strain increment $d \varepsilon$ the plastic field lose
the memory of the ``fast'' correlation between the plastic events. The
mean field thus obtained has its own dynamics on a slow
scale. Consequently, there exists a strain increment above which a
fluidity field can be defined as the slowly evolving part of the
plastic field.

\subsection{Inhomogeneous fluidity field and localization.}
Our experiment allows to follow a fluidity field which is homogeneous
at the beginning of the loading, and which condensates on a inclined
band. The fact that strain localization occurs in bi- and tri-axial
tests performed on granular materials is well known since many
decades. However, a clear understanding of this feature is still
missing~\cite{Schall2007}.  Some authors tried to relate strain
localization with properties of the fluidity-field. In a first study,
Goyon~{\it et al.}~\cite{Goyon2008} proposed a mechanism of formation
of shear band in flows of emulsion. The authors assumed that walls
that are at flow boundaries act as sources of fluidity. This fluidity
then diffuses inside the sample, forming a fluid layer close to the
wall. In Couette flow of granular material, fluidity increases
preferentially close to the rotor because of the inhomogeneity of the
stress field. In those cases, the fluidity source is localized, and
the fluidity tends to spread into the sample.

In our experiment the situation is totally opposite. Fluidity
accumulates on a band although the applied stress is increased
homogeneously, i.e. the fluidity is initially homogeneous and finally
heterogeneous. The single property that the fluidity diffuses inside
the sample is not sufficient to explain this feature. Recently, Benzi
and coworkers~\cite{Benzi2016} have shown that shear band formation
starting from an initially homogeneous fluidity in a fluidity model
should be possible.

Another possible approach to understand this localization process can
be to consider that there is some memory of where plasticity has
already occurred. In the case of cohesive materials, the introduction
of a modification of the local elastic modulus in mesoscopic models is
known to lead to some localization~\cite{Amitrano1999}. But the memory
of previous local plastic activity is not specific to cohesive
materials and can take different forms. It may described any local
structural modification due to plasticity as for example a
modification of the local structure or of the local packing fraction
as well as the wear of some frictional contacts. In the case of shear
band formation in granular materials, it is known that the packing
fraction inside the flowing band takes a particular value (the
so-called critical state)~\cite{Andreotti2013}. The damage internal
variable to take into account might then be the local packing fraction
which should evolve when plastic events take place. The introduction
of some memory or history-dependent effect is known to generate
shear-banding in mesoscopic
models~\cite{Martens2012,Vandembroucq2011}.

\subsection{Fluidity and Mohr-Coulomb theory}
As mentionned in the Introduction, the Mohr-Coulomb theory gives a
first order description of failure threshold and shear bands
orientation for frictional
materials~\cite{Nedderman1992,Andreotti2013}. This theory introduces a
frictional internal angle which is closely linked to the angle of
repose for a granular material. Localisation is predicted to occur
along a line which inclination is given by the internal friction. This
picture of failure is very simplified but captures the prominent
features observed experimentally.

In the previous parts we have shown that a shear band can be described
by an inhomogeneous concentration of fluidity. Thus, it is very
appealing to describe the emergence of a shear-band with a spontaneous
orientation in the framework of fluidity theory. To our knowledge, in
the current versions of this theory, shear bands orientation is always
given by the loading geometry. It is very likely that a theory relying
on a single internal variable is insufficient to predict the emergence
of an angle in an homogeneous stress field. As discussed in the
previous part, the introduction of another internal variable
describing a form of damage might be necessary to display localization
along an inclined line. This damage variable is likely to be linked to
the local packing fraction. Nevertheless, the existence of a
macroscopic friction in the absence of dilantancy for frictionless
particles~\cite{Peyneau2008} suggests that packing fraction might not
be the right variable.

Very recently, a mesoscopic model using a local Mohr-Coulomb criterion
and taking into account the tensorial nature of the stress
redistribution have been able to reproduce macroscopic localization in
the absence of a damage variable~\cite{Karimi2018} as well as an
orientation for avalanches different from the local coupling. Taking
into account the redistributed pressure in addition to the deviatoric
stress suggests also to introduce another internal variable to
describe the effect of the pressure redistribution.

\section{Conclusion}
In this work we have discussed the nature of the spatio-temporal
correlations observed in the plastic response of a granular material
submitted to a biaxial test. We have detailed a procedure allowing us
to separate a fluctuating component of the plastic field at small
strain increments from a slowly evolving component. Those two
components correspond to two different types of behaviors: the two
fields have independent and coexisting characteristic orientations. We
discuss the interpretation of those two behaviors in the framework of
the present debated theories of plasticity and rheology of amorphous
materials. We argue that the slowly varying component of the plastic
field is a good candidate to be interpreted as the so-called fluidity
field introduced in the last ten years to describe nonlocal effects in
the flow of amorphous materials. We underline the importance of
coarse-graining in strain increment in the definition of this field.

The question of the independence or correlation between the fast and
slow components of the field is still open. Even if the two
orientations observed are apparently unrelated it could be possible
that the slow field inherits its characteristics from the fluctuations
in a non trivial manner.

\section{Acknowledgements}
The authors thank Sean McNamara and J\'er\^ome Weiss for many
scientific discussion and acknowledge funding from Agence Nationale de
la Recherche (ANR ``Relfi'', ANR-16-CE30-0022).


\end{document}